\documentstyle[11pt]{article} 
\input epsf 
\begin{document} 
\title{Building galaxy models with Schwarzschild method and spectral dynamics}  
\author{HongSheng Zhao \thanks{Contribution to 
IAU Colloquium 172 on ``Impact of Modern Dynamics in Astronomy'', July 
6-11, 1998, Namur, Belgium; ed. S. Ferraz-Mello (Dordrecht:Kluwer)} 
\\ Sterrewacht Leiden, Niels Bohrweg 2, 
2333 CA, Leiden, NL \\(hsz@strw.LeidenUniv.nl)} 
\maketitle 
\begin{abstract} 
Tremendous progress has been made recently in modelling the morphology
and kinematics of centers of galaxies.  Increasingly realistic models
are built for central bar, bulge, nucleus and black hole of galaxies,
including our own.  The newly revived Schwarzschild method has played
a central role in these theoretical modellings.  Here I will highlight
some recent work at Leiden on extending the Schwarzschild method in a
few directions.  After a brief discussion of (i) an analytical approach
to include stochastic orbits (Zhao 1996), and (ii) the ``pendulum
effect'' of loop and boxlet orbits (Zhao, Carollo, de Zeeuw 1999), I
will concentrate on the very promising (iii) spectral dynamics method,
with which not only can one obtain semi-analytically the actions of
individual orbits as previously known, but also many other physical
quantities, such as the density in configuration space and the
line-of-sight velocity distribution of a superposition of orbits
(Copin, Zhao \& de Zeeuw 1999).  The latter method also represents a
drastic reduction of storage space for the orbit library and an
increase in accuracy over the grid-based Schwarzschild method.
\end{abstract} 
 
\section{Introduction} 
 
One of the classical problems in galaxy dynamics is building
equilibrium models for a galaxy with an observed light
distribution. The basic process can be illustrated by the simpliest
form of the problem, which is to construct a spherical, isotropic
model with a constant mass-to-light ratio $M/L$ and a stellar
distribution function $DF$ which fits the light profile.  This has the
well-known solution (Eddington 1916),
\begin{equation} 
DF(E) \propto  
\int_{E}^{0} {d^2 \rho \over d\phi^2} {d\phi \over \sqrt{\phi-E}}, 
\end{equation} 
which is a function of the energy $E$ only, 
where the potential $\phi$ comes from solving the Poisson equation, 
and the volume density $\rho(r)$ comes from  
deprojecting the light profile $\mu(R)$ 
\begin{equation} 
\rho(r)\propto {M \over L} \int_{r}^{\infty} {d \mu(R) \over dR} {dR \over \sqrt{R^2-r^2}}. 
\end{equation} 
In general  
deprojecting the light and getting the potential are relatively easier 
parts of the problem. 
 
While a simple problem in concept, it is challenging to extend the 
mathematical and numerical machinery to cope with realistic systems. 
In particular, galaxies are almost always flattened, and sometimes 
triaxial.  They are also anisotropic in velocity distribution due to 
dissipational and dissipationless processes in formation.   By 
formation they are often dominated by dark matter at very small and 
very large radii (central black holes as indicated by nuclear 
activities in AGNs and outer dark halos as by flat HI rotation 
curves).  In short, none of the three simplifying assumptions 
(constant $M/L$, isotropic and spherical) are generally valid. 
 
While progress has been made in the analytical direction, the
application is generally limited.  The Hunter \& Qian (1993) method,
for example, can construct two-integral models -- with a DF being
function of energy and angular momentum azimuthal component
$DF(E,L_z)$ -- for axisymmetric galaxies and has been applied, e.g.,
in the case of the nucleus of M32 (Qian et al. 1995.  See also,
Dejonghe 1986, Dehnen \& Gerhard 1994 for alternative techniques of
building two-integral models).  Formulaism also exists for building
anisotropic non-axisymmetric models as long as the potential remains
in St\"ackel form (Teuben 1987, Statler 1987, 1991, Arnold et
al. 1994, Dejonghe et al. 1995), and in a few cases for tumbling
models (e.g. Freeman 1966, Vandervoort 1980).  As a side comment
separability is no guarantee for self-consistency; for example, the
recent non-axisymmetric disc potentials by Sridhar \& Touma (1997)
require unphysically negative DF (Syer \& Zhao 1998).
 
At the other end of the spectrum of methods straight N-body
simulations can deal with all geometries (Aarseth \& Binney 1976,
Wilkinson \& James 1982, Barnes 1996), but their power is again limited when it
comes to sculpturing a simulation to fit a set of observations in
certain $\chi^2$ sense.  The limitation here is the huge amount of
computation to cover enough degrees of freedom to find the true best
fitting model, e.g. Fux's models (1997) for the Milky Way.

The most promising approach so far is the so-called Schwarzschild
(1979) method, after his pioneering efforts in this direction.
Basically one tries to match the observed distribution of the light
with typically a few hundred or thousand building blocks with each
being one stellar orbit populated with certain amount of stars.  One
adjusts the mass assigned to each orbit until a best match is obtained
(see reviews by Binney 1982, de Zeeuw \& Franx 1991, de Zeeuw 1996,
Merritt 1996, 1999).

\begin{figure} 
\epsfysize=6cm 
\leftline{\epsfbox[0 0 778 330]{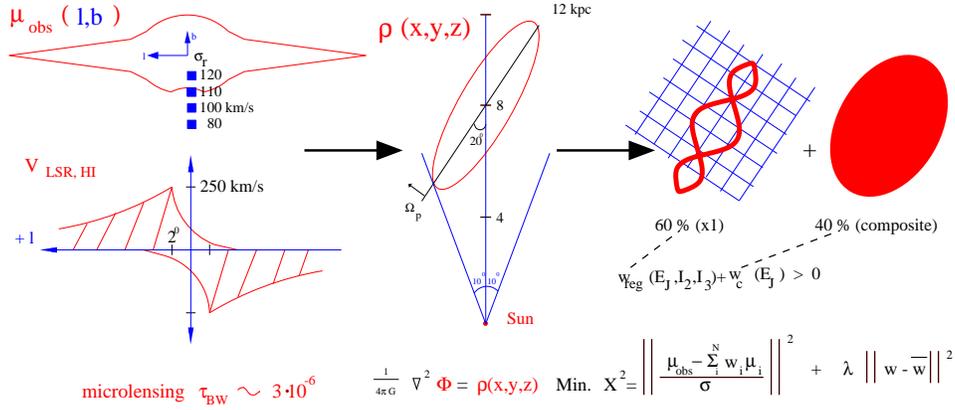}} 
\caption{An illustration of the steps in building a bar model for the Milky 
Way with a modified Schwarzschild method.  We start from (1) a 
dust-corrected near infrared surface brightness map $\mu_{obs}(l,b)$ 
of the Galaxy from COBE/DIRBE observations, (2) the velocity 
dispersion $\sigma_r$ measurements on the minor axis of the Galaxy, 
(3) the atomic hydrogen intensity map in the longitude $l$ vs. line of 
sight velocity plane, which is basically a measure of the rotation curve 
$V_{LSR,HI}$ of the Galaxy, and (4) the probability (optical depth) of 
microlensing $\tau_{BW}$ towards the Baade Window ($l=1^o,b=-4^o$). 
We then use these input quantities to build the potential 
$\Phi(x,y,z)$ of the bar via the Poisson equation, and constrain the axis 
ratio and the orientation of the bar.  Finally we seek a positive 
DF of regular orbits ($x_1$, $x_2$, $x_4$ orbits etc.) of 
weights $w(E_J,I_2,I_3)$ plus super-orbits (composite orbits) of weights 
$w(E_J)$ such that the ensemble $\sum_i^N w_i \mu_i$ matches the 
observed distribution $\mu_{obs}$ in a set of rectangular grid cells. 
The last step is a $\chi^2$ minimization problem, regularized by 
slightly ($\lambda \sim 10^{-4}$) penalizing DF where adjacent orbits 
in phase space have wildly different weights $w_i$.  We feed the final 
model to an N-body code to test stability.  }\label{mwbar.ps} 
\end{figure} 
 
\section{Schwarzschild method with bells \& whistles} 
 
The Schwarzschild method has now been extensively applied to study
nearby elliptical and S0 galaxies under the assumption of a static
spherical, axisymmetric or triaxial potential (e.g., Richstone \&
Tremaine 1988, Merritt \& Fridman 1996, Rix et al. 1997, van der Marel
et al. 1998), and has also been applied to build 2-dimensional models
of external bars (Pfenniger 1984, Wozniak \& Pfenniger 1997, see also
Sellwood \& Wilkinson 1993) and 3-dimensional models of the tumbling
bar of our own galaxy (Zhao 1996).  These applications have also
greatly generalized the original layout of the Schwarzschild method,
and in particular, it is possible to match the orbits to a variety of
kinematic data of gas and stars, and to derive a smooth physical
solution (cf. the schematic Fig.~\ref{mwbar.ps}).  Nevertheless there
are three main limitations of Schwarzschild approach and these are
best overcome by joining force with the analytical and the N-body
approaches.
 
Limitation A: Stability of a Schwarzschild model needs to be addressed 
by an N-body simulation.  An interesting idea, due to Syer \& Tremaine 
(1996) is to do the $\chi^2$ fitting and N-body simulation at the same 
time, adjusting the mass of each particle as the simulation evolves 
towards a best match of data with the distribution of the particles. 
A simpler, better understood approach is to design a Schwarzschild model 
first, then populate each library orbit with $N_{\bf A}$ particles 
with random phase where $N_{\bf A}$ is proportional to the weight 
assigned to the orbit with actions ${\bf A}$, and finally feed 
these particles to an N-body simulation code to test stability.  This 
has been applied successfully to the Galactic bar, which is found to 
be stable (Zhao 1996). 
 
Limitation B: Stochastic orbits in a Schwarzschild model make the
model evolve on time scales of the mixing time (several hundred
dynamical time, cf. Merritt \& Valluri 1996).  Merritt \& Fridman
(1996) propose to average out this effect by explicitly summing up
many stochastic orbits to achieve a good phase-mix.  This is
challenging because it means integrating a few hundred orbits for a
few thousand of dynamical times to beat down the time-dependent
fluctuations.  An alternative approach has been used in the case of
the Galactic bar (Zhao 1996).  The hybrid model makes use of two types
of building blocks for the Galaxy (cf. Fig.~\ref{mwbar.ps}): a library
of regular orbits obtained by direct integration for several hundred
dynamical time, and a library of ``super-orbits'', which are nothing
but many delta-like DFs $\sum_i N_i \delta(E_J-i \Delta)$, where the
weighting $N_i$ are to be found by the same Non-Negative Least Square
fitting code as with the weighting of the regular orbits.  Each delta
function includes all orbits with the same Jacobi integral $E_J\equiv
E-\Omega J_z$ implicitly, where $\Omega$ is the tumbling speed of the
bar and $E_J$ is the only analytical integral. Such a prescription
naturally incorporates stochastic orbits in the model without
explicitly making the division of the fraction of mass in stochastic
orbits vs. regular ones.  Variations of our analytical way of
including stochastic orbits have now been developed to model
axisymmetric systems and bars by the dynamics groups at Leiden
(Cretton et al. 1998, private communication) and Oxford (H\"afner et
al. 1998, private communication).
 
\begin{figure}  
\epsfxsize=10truecm  
\centerline{\epsfbox{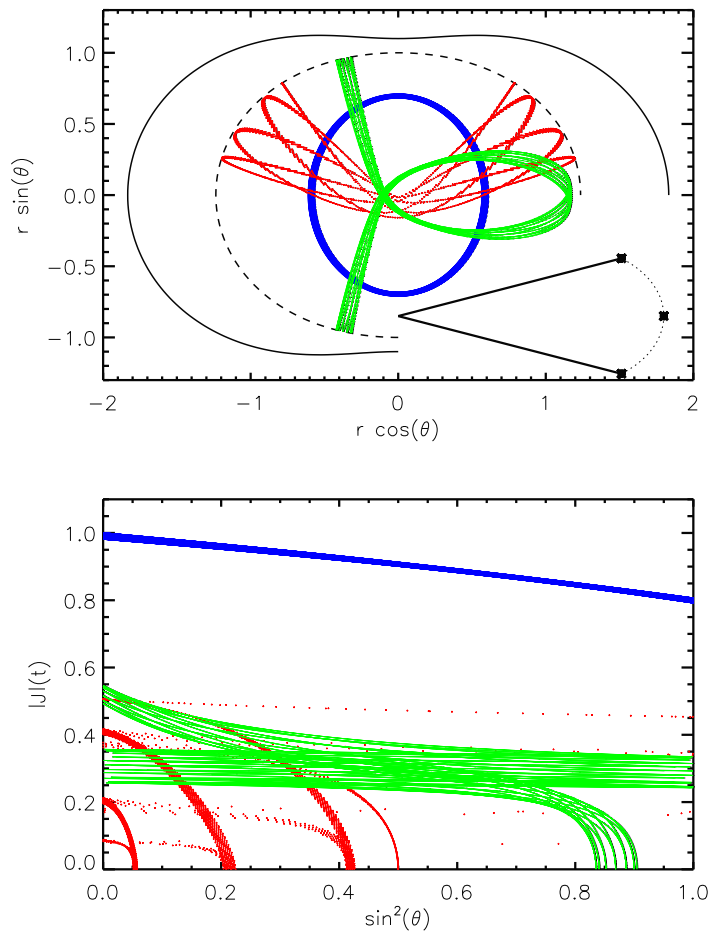}} 
\caption{ the upper panel shows three types of centro-phobic orbits: a
loop orbit (thick band), a banana orbit (dotted lines) and a fish orbit (solid
thin lines) together with the iso-density and iso-potential contours (heavy
solid and dashed lines respectively).  The lower panel shows the
angular momentum $|J(t)|$ of a star as a function of the angle from
the major axis ($\theta=0$) along the same three orbits, where each
dot is one time step of the orbit.  Note that $|J(t)|$ peaks on
approaching the major axis for all boxlet orbits, like it does for the
loops and the pendulum.  A pendulum with a variable length is also
sketched in the upper panel.  }\label{rj.ps}
\end{figure}  
Limitation C: A Schwarzschild model is cell-dependent.  Checking
self-consistency of the model involves computing the amount of time an
orbit spends inside a cell and comparing it with the amount of mass
prescribed in the same cell.  However it is possible to make
cell-independent modeling.  For example, to keep a triaxial galaxy in
equilibrium requires a healthy mix of shapes of its building blocks
with some orbits more flattened than the potential, some less
flattened.  It is well-known that loop orbits cannot reproduce a
self-consistent triaxial potential because they move too fast and
spend too little time at the major axis to match the relatively
(compared to, say, the minor axis) high model density there.  We find
that this problem is actually more general (Zhao, Carollo \& de Zeeuw
1999): it is easy to prove analytically that any regular orbit will
reach a local maximum for its angular momentum $|J(t)|$ at the major
axis, because the torque of triaxial potential is always directed
towards the major axis (cf. Fig.~\ref{rj.ps}).  So in this regard a
loop orbit or a boxlet orbit (with the shape of a banana, fish,
pretzel etc) behaves like a pendulum with a stretchable length.  Since
a pendulum tends to swing too fast and spend too little time at its
symmetry axis, the "pendulum effect" generally prevents loops and
boxlets from putting many stars at the major axis.  This can be used
as a cell-independent argument against making strongly flattened and
triaxial galactic nuclei with bananas, fishes etc., consistent with
previous authors (e.g., Gerhard \& Binney 1985, Pfenniger \& de Zeeuw
1989).
 
\section{Spectral dynamics method} 
\begin{figure}  
\epsfysize=13cm  
\centerline{\epsfbox[140 60 450 640]{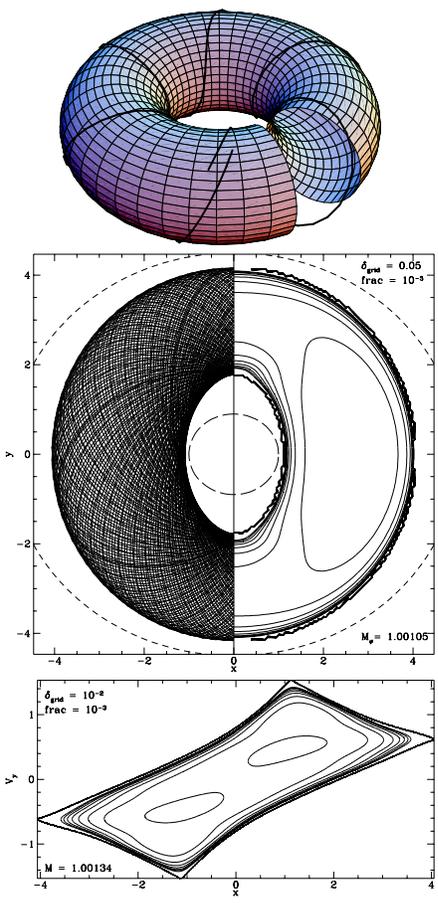}} 
\caption{ 
A planar loop orbit in various cuts of the phase space.  The top 
panel shows the phase space torus of the orbit, which will be 
populated uniformly after integrating the orbit for a long time.  The 
orbit in the configuration space is related to the torus by eq. (3), 
and the uniformly populated torus projects to a non-uniform 
distribution in the configuration space via a Jacobian (cf. eq. 6). 
The middle panel shows the reconstructed orbit in the $xy$ plane, 
folded to the left half, and the density contour map of the orbit, 
folded to the right half.  The bottom panel shows the line of sight 
velocity distribution of the orbit in the impact parameter $x$ 
vs. velocity $v_y$ plane, computed with eq. (7); the tiled 
parallelogram is indicative of the rotation of the loop (Copin et al. 1999). 
}\label{loop-torus.ps} 
\end{figure} 

Another very promising cell-independent method of building galaxies is
the spectral dynamics method.  This method, as introduced by Binney \&
Spergel (1982), provides a conceptually simple representation of a
regular orbit, by decomposing it into a truncated Fourier series
involving three fundamental frequencies.  The basic idea here is that
a regular orbit in a 3D potential is simplest described in the action
angle space since it satisfies periodic boundary conditions on the
torus (cf. Fig.~\ref{loop-torus.ps}).  Let an orbit be labeled by its
three actions ${\bf A}$, then the phase space coordinates $[{\bf
x}_{\bf A}(t),{\bf v}_{\bf A}(t)]$ are periodic with respect to the
three action angles $\theta\equiv(\omega_1,\omega_2,\omega_3) t$, ie.,
we have the following truncated Fourier series
\begin{equation}\label{xt} 
{\bf x}_{\bf A}(t) = \sum_{\lambda \equiv (l,m,n)}^{L} X_\lambda  
	\cos(\lambda\cdot\theta + \chi_\lambda),~~ 
\theta\equiv(\omega_1,\omega_2,\omega_3) t. 
\end{equation} 
where the $\omega$'s are the three basic frequencies, the coefficients 
$X_\lambda$ are the amplitudes of each frequency combination and  
$L$ is the highest order harmonics before truncation. 
Similarly the velocity of the orbit at any time, related to the position 
by a time derivative, can be written down as 
\begin{equation}\label{vt} 
{\bf v}_{\bf A}(t) = -\sum_{\lambda \equiv (l,m,n)}^{L} \omega_\lambda X_\lambda  
	\sin(\lambda\cdot\theta + \chi_\lambda), 
	~~\omega_\lambda=l \omega_1+m \omega_2 + n\omega_3, 
\end{equation} 
It is easy to work out the actions ${\bf A}$ by integrating along one of 
the three action angles, 
\begin{equation} 
{\bf A} = {1 \over 2} \sum_{\lambda} X_\lambda^2 (l \omega_1, m \omega_2,  n\omega_3). 
\end{equation} 
 
This method actually goes back to many years ago
(e.g., Ratcliff, Chang \& Schwarzschild 1984), but recent work by
the Oxford group (e.g., Kaasalainen \& Binney 1994), 
and by Papaphilipou \& Laskar (1996) and Carpintero
\& Aguilar (1998) has made it possible to extract the basic
frequencies numerically from the time series data of a regular orbit.
Namely the step from $[{\bf x}_{\bf A}(i \Delta t), {\bf v}_{\bf A}(i
\Delta t)]$ to $[\omega_\lambda, X_\lambda]$.

Most important to the Schwarzschild method is that we can compute the 
volume density of the orbit ${\bf A}$ at a given point $(x,y,z)$. 
Since we know that a regular orbit is uniformly distributed in its 
action angle space, the density in the real space is given by 
\begin{equation} 
\rho_{\bf A} (x,y,z) = {1 \over (2\pi)^3} \left| 
{\partial (x,y,z) \over \partial (\theta_1,\theta_2,\theta_3)}\right|^{-1}, 
\end{equation} 
where the partial derivatives are simply the Jacobian for the 
transformation between the action angle space 
$(\theta_1,\theta_2,\theta_3)$ and the coordinate space $(x,y,z)$. 
Since the Jacobian can be evaluated analytically with eq.~(\ref{xt}), 
we have derived a rigorous expression for the spatial distribution of 
an orbit.  Likewise the line-of-sight velocity ($v_z$) distribution of 
an orbit in the direction $(x,y)$ (cf. Fig.~\ref{loop-torus.ps}) is 
given by 
\begin{equation} 
LOSVD_{\bf A}(x,y,v_z) = {1 \over (2\pi)^3} 
	\left|{\partial (x,y,v_z) \over \partial (\theta_1,\theta_2,\theta_3)}\right|^{-1}. 
\end{equation} 
For details see Copin et al. (1999). 
 
The beauty of this method is that the description of regular orbits is
conceptually simple.  The description is time-independent, and
involves no gridding and binning.  It is also easy to store and
recover an orbit, thus saving the amount of disc space for storing
orbit libraries.  Typically the number of quantities to store is about
$10$ times the dimension of the problem; this includes the basic
frequencies and the leading amplitudes.
 
To conclude we remark that the most promising 
method might be some kind of generalized Schwarzschild method or 
hybrid method, where the computationally intentive stochastic orbits 
are implicitly modelled by the analytical super-orbits, and the 
spatial and velocity distribution of the regular orbits are treated 
with spectral dynamics. 

I thank Tim de Zeeuw for a critical reading of an earlier version and
Danny Pronk for making the electronic version of Figure 1.

{} 
 
\end{document}